# RoboBlimp: Enhancing Middle School STEM through Educational Bioinspired Blimps

Alexia M. De Costa

*College of Engineering and Computing | George Mason University*
adecosta@gmu.edu

*Abstract*— This study investigates the educational potential of *Flappy*, a low-cost, bioinspired robotic blimp platform modeled after the motion of manta rays, as a hands-on STEM learning tool for middle school students. Building on prior research emphasizing the role of social and bioinspired robotics in education, a one-day workshop was developed to introduce ten students to fundamental concepts in physics, engineering, and computer science. Participants constructed and programmed their own robotic blimps while engaging with a custom curriculum that incorporated visuals and collaborative activities. Quantitative analysis using pre- and post-assessments revealed significant learning gains, supported by a Wilcoxon Signed-Rank Test (p = 0.00195). Qualitative observations showed high levels of engagement, teamwork, and increased confidence with technical vocabulary and tools. The results suggest that affordable, bioinspired robotics platforms like *Flappy* can effectively enhance STEM comprehension and enthusiasm among younger learners, particularly when paired with structured, interactive instruction.

*Keywords*— bioinspired robotics, STEM education, robotic blimp, educational robotics

## I. Introduction

Previous research has explored the use of robotic educational agents, such as Darwin [1], Robovie with LEGO Mindstorms [2], NAO [3], and Keepon [4], to enhance student performance. These studies consistently suggest that the inclusion of social elements in robots can significantly boost student engagement and academic outcomes.

In addition to interaction-based robots, hands-on development kits like Festo's *Bionics4Education* introduce students to bioinspired robotics through creative, nature-themed designs such as robotic fish, chameleon grippers, bionic flowers, and elephant trunks [5]. Festo's kits, composed of affordable components like servo motors, plastic parts, and basic electronics, allow students to freely experiment and customize their builds using available materials. The bioinspired focus encourages learners to draw from the natural world, fostering an interdisciplinary learning experience across biology, physics, computer science, and engineering.

This study builds on these foundations by examining the effectiveness of an affordable, bioinspired robotics blimp workshop for a middle school audience. The goal was to evaluate whether the workshop could produce measurable educational benefits in this younger age group.

## II. Methods

### A. Robotic Platform

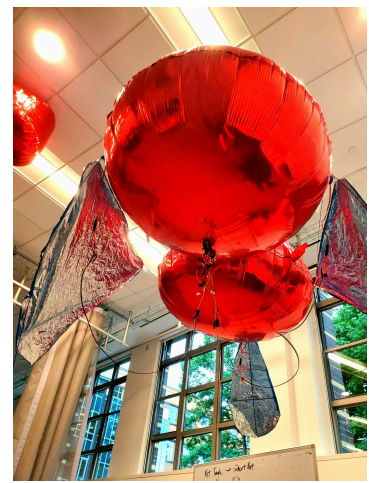

Fig. 1 *Flappy* robotic blimp.

To explore the educational impact of bioinspired robotics, this study utilized *Flappy*, an affordable robotic blimp platform modeled after the efficient undulating motion of manta rays [6]. The Flappy blimp, as shown in Fig. 1, can be assembled from materials costing approximately $100, making it an affordable option for educational environments. Its design is both a functional engineering project and a tool for introducing STEM concepts through hands-on learning.

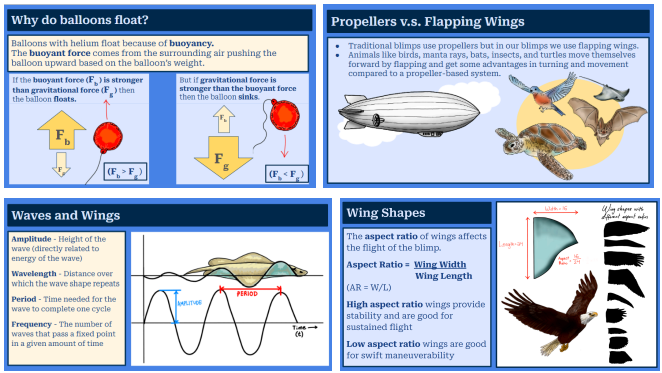

Fig. 2  Sample slides from the RoboBlimp presentation.

### B. Workshop Design

A one-day workshop was conducted with ten middle school students, divided into two teams of five. Each team was tasked with constructing and programming its own *Flappy* blimp. A custom curriculum was developed to support the workshop, featuring presentation slides, instructional photos, and visual aids to guide the students through each step of the project and explain fundamental concepts in robotics, coding, bioinspiration, aerodynamics, and engineering, as illustrated in Fig. 2.

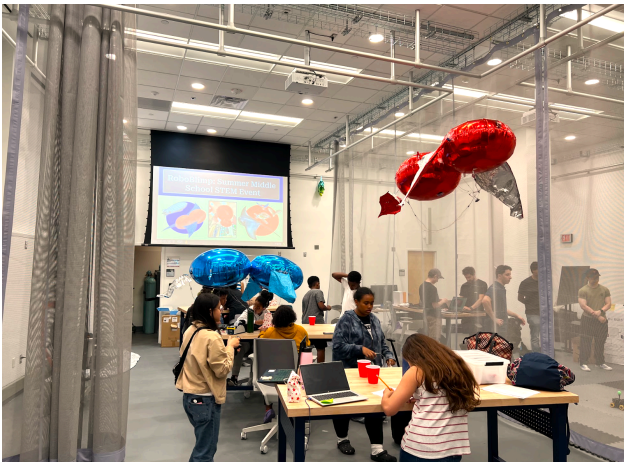
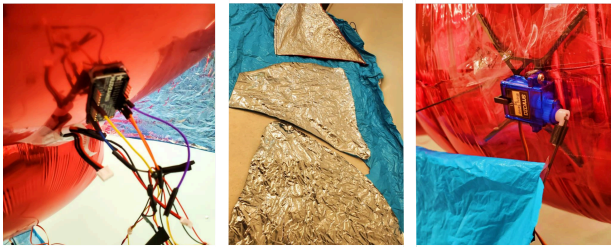

Fig. 3  Teams assembling and testing *Flappy* during the workshop.

The curriculum was structured to integrate learning objectives with engagement. Workshop activities included hands-on tasks such as:
1. Assembling the Flappy blimp (see Fig. 3),
2. Investigating fundamental physics topics such as buoyancy, center of gravity, and wave-based propulsion
3. Working with electronic components including servo motors and microcontrollers
4. Practicing wiring techniques and basic circuit construction
5. Programming the blimp using Arduino IDE, introducing basic coding

### C. Data Collection

To evaluate the effectiveness of the workshop, both quantitative and qualitative data were collected. Each participant completed a pretest before the workshop and a posttest afterward to assess changes in individual performance, comprehension, and information retention. The questions in these assessments were directly aligned with the workshop curriculum and covered key topics such as servo motor function, wiring methods, schematics, buoyancy, bioinspiration, and aerodynamics. These topics reflected the core learning objectives presented during the lessons and hands-on activities.

Qualitative observations were conducted throughout the event, documenting participant engagement, teamwork, and interaction with both the materials and concepts. These observations provided context for interpreting the test results and offered insight into how students responded to the workshop.

### III. RESULTS

#### A. Quantitative Results

Across all participants, posttest scores showed an improvement compared to pretest results, as shown in Fig. 4. The mean pretest score was 12.1 out of 46, while the mean posttest score increased to 25.9 out of 46, indicating enhanced comprehension following the workshop.

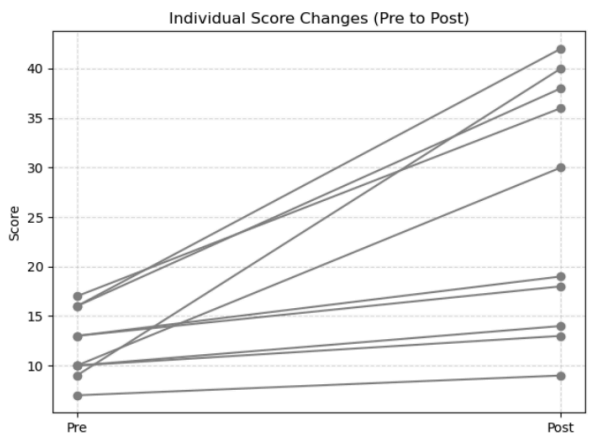

Fig. 4  Individual Score Improvements from Pretest to Posttest

*B.  Statistical Analysis*

To evaluate whether the workshop had a significant effect on student performance, a Wilcoxon Signed-Rank Test was conducted, comparing pretest and posttest scores from the sample of ten students. This non-parametric test was selected due to evidence of non-normality in the distribution of score differences as indicated by the Q-Q plot in Fig. 5.

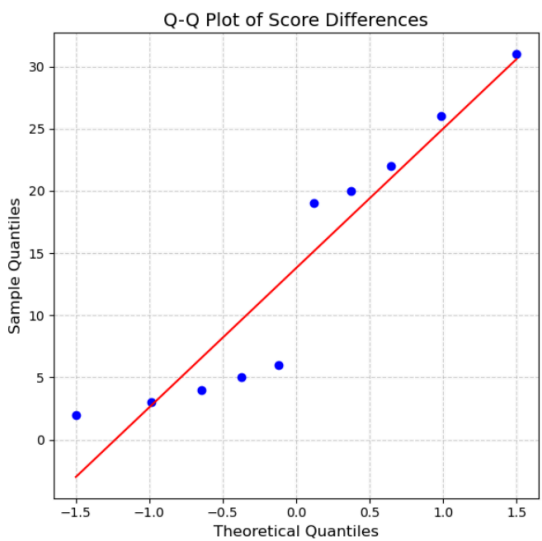

Fig. 5  Q-Q Plot of Individual Score Differences from Pretest to Posttest

The differences between posttest and pretest scores were all positive, indicating that every student improved after the workshop. The test yielded a Wilcoxon statistic of 0.0 and a p-value of 0.00195, which is well below the standard significance threshold of 0.05.

*C.  Qualitative Observations*

Throughout the workshop, students demonstrated growing confidence in both hardware and software components. Observational notes highlighted strong collaboration within both teams, with participants asking questions, assisting peers, and showing enthusiasm, particularly during the assembly, programming, and testing phases. Half the students were actively engaged, often proposing creative solutions to design challenges. The integration of visual materials appeared to enhance concept retention, as many students made direct connections between the slide content and hands-on activities. Informal conversations further revealed increasing comfort with technical vocabulary by the end of the workshop.

Notably, the five students who participated consistently and attentively in all stages of the blimp-building process exhibited the greatest improvements in their test scores. In contrast, the remaining participants, who were less engaged and at times distracted or disinterested, showed more modest gains. This difference in engagement levels likely contributed to the non-normal distribution of score differences observed in the Q-Q plot.

## IV. Conclusions

This study highlights the educational value of using affordable, bioinspired robotics platforms, like *Flappy*, as a tool for introducing middle school students to STEM concepts. The bioinspired approach not only introduces fundamental physics principles such as buoyancy, center of gravity, and wave motion, but also demonstrates how nature can serve as a powerful source of inspiration.

The hands-on construction of robotic blimps gave students practical experience in engineering design, electronics, and basic programming. Working collaboratively to build and test their blimps encouraged teamwork, problem-solving, and iterative learning. This environment proved to be both educational and engaging, particularly for students who actively participated throughout the workshop.

Importantly, the workshop showed that low-cost educational robotics platforms can effectively support skill development. Students gained exposure to technical tools and vocabulary while developing soft skills such as communication, persistence, and creativity. The success of this one-day event suggests that active engagement, combined with a well-structured, interdisciplinary curriculum, can make a meaningful impact on student learning and enthusiasm for STEM.

ACKNOWLEDGMENT

I would like to express my sincere gratitude to Dr. Daigo Shishika and Dr. Peter Plavchan for their mentorship and guidance throughout this project. I am also deeply thankful to Professor Kerin Anne Hilker-Balkissoon for her generous support through the Research & Interdisciplinary STEM Experiences (RISE) Scholar Initiative. Special thanks to Kentaro Nojima-Schmunk for his insightful advice and hands-on assistance during the workshop, which contributed greatly to its success.

REFERENCES

[1] L. Brown, R. Kerwin and A. M. Howard, "Applying Behavioral Strategies for Student Engagement Using a Robotic Educational Agent," *2013 IEEE International Conference on Systems, Man, and Cybernetics*, Manchester, UK, 2013, pp. 4360-4365, doi: 10.1109/SMC.2013.744.

[2] T. Kanda, M. Shimada, and S. Koizumi. 2012. "Children learning with a social robot." In Proceedings of the seventh annual ACM/IEEE international conference on Human-Robot Interaction (HRI '12). Association for Computing Machinery, New York, NY, USA, 351–358. https://doi.org/10.1145/2157689.2157809

[3] E. A. Konijn, J. F. Hoorn, "Robot tutor and pupils' educational ability: Teaching the times tables," *Computers & Education*, vol. 157, p. 103970, 2020. [Online]. Available: http://www.sciencedirect.com/science/article/pii/S0360131520301688

[4] D. Leyzberg, S. Spaulding, and B. Scassellati. 2014. "Personalizing robot tutors to individuals' learning differences." In Proceedings of the 2014 ACM/IEEE international conference on Human-robot interaction (HRI '14). Association for Computing Machinery, New York, NY, USA, 423–430. https://doi.org/10.1145/2559636.2559671

[5] Festo. 2017. Bionics4Education | Festo USA. https://www.festo.com/us/en/e/about-festo/research-and-development/bionic-learning-network/highlights-from-2015-to-2017/bionics4education-id_32696/

[6] K. Nojima-Schmunk et al., "Manta Ray Inspired Flapping-Wing Blimp," *2024 IEEE International Conference on Robotics and Biomimetics (ROBIO)*, pp. 2166–2173, Dec. 2024, doi: https://doi.org/10.1109/robio64047.2024.10907498.